\documentstyle[12pt,aaspp]{article}

\received{29 June 1993}
\accepted{24 February 1994}
\journalid{431}{20 August 1994}

\slugcomment{ApJ, 431, in press (20 August 1994)}

\begin{document}

\title{Gravitational Instability of Cold Matter}

\author{Edmund Bertschinger and Bhuvnesh Jain}
\affil{Department of Physics, MIT, Cambridge, MA 02139}

\begin{abstract}
We solve the nonlinear evolution of pressureless, irrotational density
fluctuations in a perturbed Robertson-Walker spacetime using a new
Lagrangian method based on the velocity gradient and gravity gradient
tensors.  Borrowing results from general relativity, we obtain a set
of Newtonian ordinary differential equations for these quantities
following a given mass element.  Using these Lagrangian fluid equations
we prove the following collapse theorem:
A mass element whose density exceeds the cosmic mean at high redshift
collapses to infinite density at least as fast as a uniform spherical
perturbation with the same initial density and velocity divergence.
Velocity shear invariably speeds collapse --- the spherical tophat
perturbation, having zero shear, is the slowest configuration to collapse
for a given initial density and growth rate.  Two corollaries follow:
(1) Initial density maxima are not generally the sites where collapse
first occurs.  The initial velocity shear (or tidal gravity field) also
is important in determining the collapse time.
(2) Initially underdense regions undergo collapse if the shear is
sufficiently large.

If the magnetic part of the Weyl tensor vanishes, the nonlinear
evolution is described purely locally by these equations.  This condition
is exact for highly symmetrical perturbations (e.g., with planar,
cylindrical, or spherical symmetry) and may be a good approximation
in many other circumstances.  Assuming the vanishing of the magnetic part
of the Weyl tensor we compute the exact nonlinear gravitational evolution
of cold matter.  We find that 56\% of initially underdense regions
collapse in an Einstein-de Sitter universe for a homogeneous and
isotropic random field.  We also show that, given this assumption,
the final stage of collapse is generically two-dimensional, leading to
strongly prolate filaments rather than Zel'dovich pancakes.  While this
result may explain the prevalence of filamentary collapses in some N-body
simulations, it is not true in general, suggesting that the magnetic
part of the Weyl tensor does not necessarily vanish in the Newtonian limit.
\end{abstract}

\keywords{cosmology: theory --- large-scale structure of the universe
--- gravitation}

\section{Introduction}

According to the gravitational instability theory, cosmic structure
developed from the amplification of small density fluctuations generated
in the early universe.  During the linear stage the amplitude is small
and the evolution is described simply by the linearized equations of
motion (gravitational field equations plus Boltzmann or fluid equations
for the matter and radiation components).  If pressure gradients are not
too large, self-gravitating fluctuations eventually become nonlinear.
The difficulty of following the nonlinear evolution analytically has led
cosmologists to resort to gravitational N-body simulations for studying
the formation of galaxies and large-scale structure (e.g., \cite{B91}
and references therein).  However, these simulations should be accompanied
by analytical models enabling one to check and comprehend the results.

There are two nonlinear models that have been widely used to describe
nonlinear gravitational evolution in cosmology.  The first is the
spherical model, for which a simple exact solution exists (\cite{To34};
\cite{Bo47}; \cite{PP67}; \cite{GG72}; \cite{P80}, \S 19).  For
pressureless spherical collapse one follows the trajectories of spherical
mass shells having fixed enclosed mass.  This model is generally applied
to the formation of dense objects, the logic being that the first
structures to collapse are high density peaks, with a nearly spherical
distribution of matter surrounding them (\cite{BBKS86}).

The second model is the kinematical one of Zel'dovich (1970).  The key
idea here is to extrapolate the linear evolution of peculiar velocity
beyond the linear regime (for a review see \cite{SZ89}).  The Zel'dovich
method is also Lagrangian but is not restricted to spherical symmetry.
Indeed, for plane-parallel perturbations (but not for two- or
three-dimensional ones) it is exact for particles whose trajectories
have not intersected others.

The Zel'dovich approximation (as it is known in multi-dimensions) has
provided a great deal of insight into the initial nonlinear evolution
of density fluctuations.  Recently it has inspired a number of developments
concerning the evolution of the velocity field for irrotational flows.
Nusser et al. (1991) showed how to compute the density from the Eulerian
velocity field in the Zel'dovich approximation; their work was extended
to second order in Lagrangian perturbation theory by Gramann (1993).
Lagrangian perturbation theory has also been developed and applied by
Moutarde et al. (1991) and Bouchet et al. (1992).  A broader perturbative
framework has been developed by Giavalisco et al. (1993).  The Zel'dovich
approximation has also been used to estimate the probability distribution
of density in the nonlinear regime (\cite{K91}; \cite{PS93}; \cite{KBGND94}).
Bernardeau (1992) has used a somewhat different statistical summation method
to relate the density and velocity divergence in the nonlinear regime.

The limitations of the spherical model and Zel'dovich approximation are
well known.  The first applies only with spherical symmetry while the
second gives inaccurate trajectories beyond linear perturbation theory.
Various authors have therefore suggested modifications of the Lagrangian
approach to increase its accuracy.  Buchert (1992) has investigated the
linear and nonlinear limits of the Zel'dovich approximation.  He and
Barrow \& Saich (1993) have also studied the effects of vorticity.
Gurbatov, Saichev, \& Shandarin (1980) and Kofman, Pogosyan, \& Shandarin
(1990) have proposed adding viscosity to prevent the intersection of fluid
trajectories.  Lachi\`eze-Rey (1993a,b) has recently developed elegant
matrix methods for extending the Zel'dovich approximation beyond linear
order by following the evolution of the deformation tensor, the Jacobian
of the transformation from Lagrangian to Eulerian coordinates.  Matarrese
et al. (1992) have suggested the interesting approximation of freezing
the initial fluid streamlines using the linear velocity field, in effect
treating the velocity potential as constant.  Brainerd, Scherrer, \&
Villumsen (1993) and Bagla \& Padmanabhan (1994)
have suggested instead treating the gravitational potential as constant.
While these approaches have provided valuable insights about nonlinear
evolution, they are all based on approximations to the evolution of the
gravitational field.

General relativity can contribute to a solution by providing evolution
equations for the gravitational field as opposed to the nonlocal static
Poisson equation underlying the Newtonian approach.  It has long been
known among relativists that the Einstein equations and the Bianchi
identities provide both dynamical equations and initial-value constraints
for the gravitational field, whether specified by the metric or by
the Ricci and Weyl tensors (see \cite{E71} for an excellent overview).
Moreover, the relativistic equations naturally lead to a closed set of
equations for the fluid variables and gravitational fields specified
by the Weyl tensor.  The Weyl tensor may be decomposed into electric
and magnetic parts; in the Newtonian limit the electric part represents
the gravitational tidal field.  The Bianchi identities provide an exact
Lagrangian evolution equation for the Weyl tensor and, therefore, for
the tidal field in the Newtonian limit.  Remarkably, when the magnetic
part of the Weyl tensor vanishes the evolution of the tidal field is
purely local, allowing it to be integrated together with the density
and the velocity gradient tensor independently for each mass element.

The fully relativistic Lagrangian equations of motion have been used
to describe the linear evolution of cosmological perturbations (\cite{H66};
\cite{EB89}; \cite{HV90}; Bruni, Dunsby, \& Ellis 1992).  Their application
to nonlinear evolution of cosmological fluctuations was first made by
Barnes \& Rowlingson (1989).  A major step applying these methods to
large-scale structure was made by Matarrese, Pantano, \& Saez (1993),
who showed that exact solutions exist prior to orbit-crossing for spherical
and plane-parallel perturbations, and that these solutions are identical
to the afore-mentioned spherical model and Zel'dovich solution.  They also
demonstrated a local mapping between Lagrangian and Eulerian coordinates.
The Lagrangian fluid method in general relativity has been developed further
by Croudace et al. (1993), who introduced an approximation method based on
the Hamilton-Jacobi equation.

The current paper presents the Lagrangian fluid approach in its Newtonian
framework and applies it to the general problem of nonlinear gravitational
collapse beginning from small perturbations of a Friedmann-Robertson-Walker
spacetime.  In \S 2 we present the Lagrangian fluid equations and their
solution in second-order perturbation theory.  Although these equations
have been given by Matarrese et al. (1993), our approach emphasizes the
Newtonian interpretation in contrast with their general relativistic one.
In \S 3 we present a collapse theorem giving a lower bound on the collapse
time of an overdense perturbation.  The full equations are integrated
assuming vanishing magnetic part of the Weyl tensor in \S 4, where we
present results for both overdense and underdense initial perturbations.
Conclusions are given in \S 5.

\section{Nonlinear Evolution of Density, Velocity Gradients, and Tides}

The evolution of a pressureless fluid may be described either by the
Euler equations, by geodesic equations for individual mass elements,
or by Lagrangian fluid equations.  Although our goal is a closed system
of Lagrangian fluid equations, we begin with the nonrelativistic
Eulerian approach because of its greater familiarity.  In a perturbed
Robertson-Walker universe with expansion scale factor $a(\tau)$ the Euler
equations take the following form (\cite{B92}):
\begin{equation}
{\partial\delta\over\partial\tau}+\vec\nabla\cdot\left[(1+\delta)\vec v\,
\right]=0\ ,
\eqnum{1a}
\label{eulcont}
\end{equation}
\begin{equation}
{\partial\vec v\over\partial\tau}+\left(\vec v\cdot\vec\nabla\right)\,
\vec v=-{\dot a\over a}\,\vec v-\vec\nabla\phi\ ,
\eqnum{1b}
\label{euler}
\end{equation}
\begin{equation}
\nabla^2\phi=4\pi Ga^2\delta\rho={3\over2}\,\Omega_0 H_0^2 a^{-1}\delta\ .
\eqnum{1c}
\label{poisson}
\end{equation}
\setcounter{equation}{1}
The mass density is $\rho=\bar\rho+\delta\rho=\bar\rho(\tau)(1+\delta)$
and the peculiar velocity is $\vec v=d\vec x/d\tau$ where $\tau$ is
conformal time ($d\tau=dt/a$, with dots denoting derivatives with respect
to $\tau$) and $\vec x$ is the comoving position with corresponding
gradient $\vec\nabla=\vec e^{\,i}\nabla_i=\vec e^{\,i}\partial/\partial
x^i$ with orthonormal basis vectors: $\vec e_i\cdot\vec e_j=\delta_{ij}$.
Throughout this paper we assume the background space to be flat on scales
of interest.  Equations (1) are valid for a nonrelativistic pressureless
fluid on scales much smaller than the Hubble distance $cH^{-1}$.  In adopting
a Newtonian approach we neglect gravitational waves and gravomagnetism
(the gravitational effect of vorticity).

We will rewrite the fluid equations in terms of the Lagrangian time
derivative following a fixed fluid element, $d/d\tau=\partial/\partial
\tau+\vec v\cdot\vec\nabla$.  Each Lagrangian fluid element has a trajectory
$\vec x(\vec q,\tau)$ where $\vec q$ is a Lagrangian coordinate so that
$d/d\tau$ is the time derivative at fixed $\vec q$ while $\partial/
\partial\tau$ is taken at fixed $\vec x$.  Converting the time derivatives
from Eulerian to Lagrangian, equation (\ref{eulcont}) becomes
\begin{equation}
{d\delta\over d\tau}+(1+\delta)\,\theta=0\ ,\quad
\theta\equiv\vec\nabla\cdot\vec v\ .
\label{lagcont}
\end{equation}

 From equation (\ref{lagcont}) we see that we require an evolution
equation for the velocity divergence (or expansion scalar).  For this
purpose we consider the velocity gradient (or rate-of-strain) tensor.
We decompose this tensor into its trace and traceless symmetric and
antisymmetric parts in the usual way:
\begin{equation}
\nabla_iv_j={1\over3}\,\theta\,\delta_{ij}+\sigma_{ij}+\omega_{ij}\ ,
\quad \sigma_{ij}=\sigma_{ji}\ ,\quad
\omega_{ij}=\epsilon_{ijk}\,\omega^k=-\omega_{ji}\ ,
\label{gradv}
\end{equation}
where $2\,\vec\omega=\vec\nabla\times\vec v$ is the vorticity and
$\epsilon_{ijk}$ is the three-dimensional Levi-Civita symbol.  The
unperturbed Hubble flow is absent from $\theta$ because $\vec v$
is the peculiar velocity.  Taking the divergence of equation (\ref{euler}),
using equations (\ref{poisson}) and (\ref{gradv}), and converting the
time derivative from Eulerian to Lagrangian, one obtains the Raychaudhuri
equation:
\begin{equation}
{d\theta\over d\tau}+{\dot a\over a}\,\theta+{1\over3}\,\theta^2+
\sigma^{ij}\sigma_{ij}-2\omega^2=-4\pi G\bar\rho a^2\delta\ ,
\label{raych}
\end{equation}
where $\omega^2\equiv\omega^i\omega_i$.

Next we require evolution equations for the shear $\sigma_{ij}$ and
vorticity.  These follow from the traceless symmetric and antisymmetric
parts of the gradient of equation (\ref{euler}).  The curl gives
\begin{equation}
{d\omega^i\over d\tau}+{\dot a\over a}\,\omega^i+{2\over3}\,\theta\,
\omega^i-\sigma^i_{\ j}\,\omega^j=0\ .
\label{vort}
\end{equation}
The traceless symmetric part of the gradient of equation (\ref{euler}) gives
\begin{equation}
{d\sigma_{ij}\over d\tau}+{\dot a\over a}\,\sigma_{ij}+{2\over3}\,
\theta\,\sigma_{ij}+\sigma_{ik}\sigma^k_{\ j}+\omega_i\omega_j-
{1\over3}\,\delta_{ij}\left(\sigma^{kl}\sigma_{kl}+\omega^2\right)=
-E_{ij}\ ,
\label{shear}
\end{equation}
where $E_{ij}\equiv\nabla_i\nabla_j\phi-(1/3)\,\delta_{ij}\,\nabla^2\phi$
is the gravitational tidal field, known in general relativity as the
electric part of the Weyl tensor in the fluid frame (\cite{H66};
\cite{E71}).

Equations (\ref{lagcont}) and (\ref{raych})--(\ref{shear}) give a nearly
closed system of equations for the mass density and the velocity gradient
(represented by the expansion, shear, and vorticity) following a Lagrangian
fluid element.  However, the gravitational tidal field remains.  In the
Newtonian framework, $E_{ij}$ is found after solving the Poisson equation
for $\phi$.  Because this is a nonlocal operation involving the mass
distribution everywhere, it would appear to be impossible to obtain a
set of Lagrangian equations for one mass element.

This conclusion is false.  In general relativity, the field equations
are equivalent to a set of evolution equations plus constraints.  The
Poisson equation corresponds to one of the constraints, the ADM energy
constraint equation.  However, there exist evolution and constraint
equations for all of the gravitational fields, including the tidal
tensor $E_{ij}$.  It is possible to obtain an evolution equation for
$E_{ij}$.  However, this equation is not necessarily local.

Ellis (1971) gives a clear discussion of the Lagrangian evolution
equations and constraints for gravitational fields in general relativity.
Equations of motion for the Weyl tensor (giving that part of the
gravitational field not produced by local sources) follow from the
Bianchi identities rather than the Einstein equations.  Assuming that
the matter has nonrelativistic peculiar velocities and pressure, the
evolution equation for the electric part of the Weyl tensor (\cite{E71};
\cite{MPS93}; \cite{CPSS93}), expressed in comoving coordinates, is
\begin{equation}
{dE_{ij}\over d\tau}+{\dot a\over a}\,E_{ij}+\theta E_{ij}+
\delta_{ij}\,\sigma^{kl}E_{kl}-3\sigma^k_{\ \,(i}E_{j)k}
+\epsilon^{kl}_{\ \ \,(i}E_{j)k}\omega_l
-\nabla_k\,\epsilon^{kl}_{\ \ \,(i}H_{j)l}=
-4\pi G\rho a^2\sigma_{ij}\ .
\label{tide}
\end{equation}
Parentheses around a pair of subscripts indicates symmetrization:
$\sigma^k_{\ \,(i}E_{j)k}\equiv{1\over2}\,(\sigma^k_{\ \,i}E_{jk}+
\sigma^k_{\ \,j}E_{ik})$.  The fully antisymmetric Levi-Civita tensor
is $\epsilon_{ijk}$ (with $\epsilon_{123}=+1$ in Cartesian coordinates).
The magnetic part of the Weyl tensor is denoted $H_{ij}$.  We have not
assumed that density fluctuations or the velocity gradient are small;
the total mass density $\rho$ (as opposed to the fluctuation
$\delta\rho$) appears on the right-hand side.  As in the preceding
equations, the time derivative is Lagrangian, so that each fluid element
$\vec q$ is assigned a value $E_{ij}(\vec q,\tau)$.

Aside from the term involving $H_{ij}$, equation (\ref{tide}) is
purely local.  Thus, if $H_{ij}=0$, we have obtained a closed set of
local Lagrangian equations for the nonlinear evolution of pressureless
irrotational matter.  This fact was first noted by Barnes \& Rowlingson
(1989) and applied in cosmology by Matarrese et al. (1993, 1994).  If
$H_{ij}=0$ and if equation (\ref{poisson}) is applied as a constraint
at some initial time, then integrating equation (\ref{tide}) for $E_{ij}$
gives exactly the same results (prior to orbit-crossing) as if $E_{ij}$
were computed from the solution of the Poisson equation at the final time.

Do local evolution equations for the gravitational tidal field violate
causality?  No, because they follow directly from general relativity,
which is manifestly causal.  The resolution of this paradox is the fact
that on account of the Bianchi identities the Einstein field equations
contain information about the evolution of the matter: the contracted
Bianchi identities imply local energy-momentum conservation.  Causality
is preserved as long as the motion of the matter is such as to generate
no gravitational radiation.

We shall not write down the evolution equation for $H_{ij}$ here; it
is given by Ellis (1971).  It is clear that if the gradient of $H_{ij}$
(which appears in eq. [\ref{tide}]) is nonzero, local evolution breaks
down and the evolution of the tide and velocity variables becomes much
more complicated.  In this paper we shall either restrict ourselves to
proving results that are valid independently of the value of $H_{ij}$,
or we shall assume that it can be neglected.  While the latter choice
is not entirely satisfactory, it may be regarded as an approximation
whose validity is to be established later.  By making this approximation
we can easily integrate the evolution equations and compare the results
against N-body or other fully numerical solutions.  Thus, we can make
strong predictions about gravitational collapse that can be used to
assess the importance of the magnetic part of the Weyl tensor even
if we lack a good Newtonian interpretation.

Equation (\ref{vort}) shows that if $\vec\omega=0$ initially, it remains
zero.  This statement remains valid for individual collisionless mass
elements even after trajectories intersect.  In the following we assume
that primeval vorticity is negligible so that $\vec\omega=0$.

We will see below that for growing-mode fluctuations in the linear regime
with $H_{ij}=0$, $E_{ij}\propto\sigma_{ij}$ and therefore the eigenvectors
of these matrices are aligned.  If we choose our coordinate axes to be
aligned with these eigenvectors initially, then equations (\ref{shear})
and (\ref{tide}) are diagonal, implying that the eigenvectors remain fixed
in direction (\cite{BR89}).  Thus, each of the tensor equations
(\ref{shear}) and (\ref{tide}) may be reduced to equations of motion
for two of the eigenvalues (with the sum of eigenvalues vanishing for
a traceless matrix).  For convenience we write the tensors in the following
form:
\begin{equation}
\sigma_{ij}={2\over3}\,\sigma\,Q_{ij}(\alpha)\ ,\quad
E_{ij}={8\pi\over3}\,G\bar\rho a^2\,\epsilon\,(1+\delta)\,
Q_{ij}(\beta)\ ,
\label{stq}
\end{equation}
where we have introduced new scalars $\sigma\le0$, $\epsilon\ge0$,
$\alpha$, and $\beta$, and a one-parameter traceless quadrupole matrix
\begin{equation}
[Q_{ij}(\alpha)]\equiv{\rm diag}\left[
\cos\left(\alpha+2\pi\over3\right),\,\cos\left(\alpha-2\pi\over3\right),\,
\cos\left(\alpha\over3\right)\,\right]\ .
\label{quad1}
\end{equation}
In the following we will refer to $\sigma$ and $\epsilon$ as the
shear and tide scalars, respectively.  Similarly, $\alpha$ and $\beta$
are called the shear and tide angles, respectively.  These angles give
the ratios of eigenvalues (or axis ratios) of the shear and tidal tensors.

This matrix representation is convenient because all possible eigenvalues
are obtained by $q\,Q_{ij}(\alpha)$ with $q\in[0,\infty)$ (or $-q$ in the
same range) and $\alpha\in[0,\pi]$.  Although $Q_{ij}(\alpha)$ is periodic
with period $6\pi$ rather than $2\pi$, $Q_{ij}(\alpha\pm 2\pi)$ and
$Q_{ij}(-\alpha)$ differ from $Q_{ij}(\alpha)$ only by permutation of
the coordinates.  The reduction of the angular range by a factor of six
corresponds to ordering the eigenvalues so that $Q_{11}\le Q_{22}\le
Q_{33}$.  Two of the eigenvalues are negative for $\alpha\in[0,\pi/2)$
while two are positive for $\alpha\in(\pi/2,\pi]$.  The sum of the
eigenvalues vanishes.  The square and determinant of $Q_{ij}$ are simple:
$Q^{ij}Q_{ij}={3\over2}$, ${\rm det}\,[Q_{ij}(\alpha)]={1\over4}\cos
\alpha$.  We will also need the following differential and product identities:
\begin{equation}
d\,Q_{ij}(\alpha)={1\over3}\,Q_{ij}\left(\alpha+{3\pi\over2}\right)
\,d\alpha\ ,\quad
2Q_{ik}(\alpha)Q_{kj}(\beta)=\cos\left(\alpha-\beta\over3\right)\,
\delta_{ij}+Q_{ij}(-\alpha-\beta)\ .
\label{quad2}
\end{equation}

Substituting equations (\ref{stq}) into equations (\ref{shear}) and
(\ref{tide}) and using equations (\ref{lagcont}), (\ref{quad1}), and
(\ref{quad2}), we get the following equations of motion for the shear
and tide scalars and angles:
\begin{equation}
{d\sigma\over d\tau}+{\dot a\over a}\,\sigma+{1\over3}\,\sigma\left(
2\theta+\sigma\cos\alpha\right)=-4\pi G\bar\rho a^2\,\epsilon\,
(1+\delta)\cos\left(\alpha-\beta\over3\right)\ ,
\label{dsig}
\end{equation}
\begin{equation}
{d\alpha\over d\tau}-\sigma\sin\alpha=12\pi G\bar\rho a^2\,
{\epsilon\,(1+\delta)\over\sigma}\,\sin\left(\alpha-\beta\over3\right)\ ,
\label{dalp}
\end{equation}
\begin{equation}
{d\epsilon\over d\tau}-\sigma\epsilon\,\cos\left(\alpha+2\beta
\over3\right)=-\sigma\,\cos\left(\alpha-\beta\over3\right)\ ,
\label{deps}
\end{equation}
\begin{equation}
{d\beta\over d\tau}+3\sigma\sin\left(\alpha+2\beta\over3\right)=
-{3\sigma\over\epsilon}\,\sin\left(\alpha-\beta\over3\right)\ .
\label{dbet}
\end{equation}
Equations (\ref{deps}) and (\ref{dbet}) have assumed $H_{ij}=0$.
Equations (\ref{dsig})--(\ref{dbet}) have singular solutions with
$\delta$, $-\theta$, $-\sigma$, and $\epsilon$ diverging for finite
$\tau$.  When approaching a singularity by numerical integration we
change the independent variable from $\tau$ to $x\equiv\ln(1+\delta)$ with
$dx=-\theta d\tau$ and then integrate $\tau(x)$ and $(\theta/\delta)(x)$,
$(\sigma/\theta)(x)$, and $(\epsilon/\delta)(x)$ in place of
$\delta(\tau)$, $\theta(\tau)$, $\sigma(\tau)$, and $\epsilon(\tau)$.

With zero vorticity and pressure, equations (\ref{lagcont}), (\ref{raych})
(with $\sigma^{ij}\sigma_{ij}$ becoming ${2\over3}\sigma^2$), and
(\ref{dsig})--(\ref{dbet}) completely determine the evolution of the
density, expansion, shear, and tides until fluid trajectories intersect.
We assume that at early times the perturbations were small, $\delta^2\ll1$,
and $\Omega\approx1$ (this is generally valid even if $\Omega\ne1$ today).
The solution of the linearized equations then gives $\delta=\delta_0a(\tau)$
and $\theta=-\dot\delta$.  However, we must also specify the initial
eigenvalues of the shear and tidal tensors.  For gravitationally-induced
linear perturbations these two tensors are proportional to each other,
as one may see by linearizing equation (6).  Thus, altogether we need three
constants to specify a growing-mode solution.  We take these to be the
the linear density fluctuation $\delta_0$, tide (or shear) scalar
$\epsilon_0$, and tide (or shear) angle $\alpha_0$.  These quantities
are simply related to the double gradient of the gravitational potential:
\begin{equation}
\nabla_i\nabla_j\phi=\Omega_0\,H_0^2\,\left[{1\over2}\,\delta_0\,\delta_{ij}
+\epsilon_0\,Q_{ij}(\alpha_0)\right]\ ,
\label{gradg}
\end{equation}
where we have assumed that coordinate axes have been oriented locally to
coincide with the eigenvectors of $\nabla_i\nabla_j\phi$.

In terms of the perturbation constants $(\delta_0,\epsilon_0,\alpha_0)$
the second-order solution for small $a$ is
\begin{equation}
\delta=a\delta_0+{1\over2}a^2(\delta_0^2+\theta_1)\ ,\quad
\theta=-\dot a\,(\delta_0+a\theta_1)\ ,\quad
\theta_1={1\over21}\left(13\,\delta_0^2+8\epsilon_0^2\right)\ ,
\eqnum{16a}
\label{ic1}
\end{equation}
\begin{equation}
\sigma=-\dot a\,(\epsilon_0+a\sigma_1)\ ,\quad
\sigma_1={1\over21}\left(26\epsilon_0\delta_0-5\epsilon_0^2\cos\alpha_0
\right)\ ,\quad \alpha=\alpha_0+{5\over7}\,a\epsilon_0\sin\alpha_0\ ,
\eqnum{16b}
\label{ic2}
\end{equation}
\begin{equation}
\epsilon=a(\epsilon_0+a\epsilon_1)\ ,\quad
\epsilon_1={13\over21}\left(\epsilon_0\delta_0-\epsilon_0^2\cos\alpha_0
\right)\ ,\quad \beta=\alpha_0+{13\over7}\,a\epsilon_0\sin\alpha_0\ .
\eqnum{16c}
\label{ic3}
\end{equation}
\setcounter{equation}{16}

In the linear regime the expansion scalar is proportional to minus the
density perturbation and the shear tensor is proportional to minus the
tidal tensor, implying $\nabla_i v_j\propto-\nabla_i\nabla_j\phi$.  This
proportionality occurs because for growing-mode perturbations, the peculiar
velocity is proportional to the gravity vector $-\vec\nabla\phi$.  Thus,
the initial conditions may be specified by the eigenvalues of either
$\nabla_i\nabla_j\phi$ or $-\nabla_i v_j$, as they are equivalent.  However,
the proportionality of the shear and tide is broken in second-order
perturbation theory.

\section{Collapse Theorem}

We note from equations (\ref{lagcont}) and (\ref{raych}) that shear
(nonzero $\sigma$) increases the rate of growth of the convergence
$-\theta$, thereby increasing the rate of growth of density fluctuations.
Vorticity (nonzero $\omega$) inhibits this process.  With zero shear and
vorticity the velocity gradient tensor is isotropic (eq. [\ref{gradv}]),
corresponding to uniform spherical collapse with radial motions toward
some center.  This suggests that uniform spherical perturbations grow more
slowly than others.

To extend this argument into a theorem we combine equations (\ref{lagcont})
and (\ref{raych}) into a second-order differential equation for $\delta(\tau)$:
\begin{equation}
\ddot\delta+{\dot a\over a}\,\dot\delta={4\over3}\,{\dot\delta^2\over1+
\delta}+(1+\delta)\left({2\over3}\,\sigma^2-2\omega^2+4\pi G\bar\rho a^2
\delta\right)\ .
\label{ddotdel}
\end{equation}
This equation is exact for pressureless matter, regardless of whether
the magnetic part of the Weyl tensor vanishes.  We see that if $\omega=0$,
the growth rate of $\delta$ is minimized for $\sigma=0$.  This conclusion
holds independently of assumptions about the evolution of the shear or
tides and is due to the simple geometrical fact that shear increases
the rate of growth of the convergence of fluid streamlines.

When $\sigma=\omega=0$, equation (\ref{ddotdel}) reduces to the exact
equation for the evolution of the mean density in the spherical model.
In other words, this equation is equivalent to the equation of motion
for a spherical shell of proper radius $r(t)$ enclosing a fixed mass $M$
related to $\delta$ by $1+\delta=3M/(4\pi\bar\rho r^3)$.  From the exact
solution it is known that spherically symmetric growing density fluctuations
collapse if $\delta_i=a_i\delta_0>(3/5)(\Omega_i^{-1}-1)$ where
$\delta_i$ and $\Omega_i$ refer to the density fluctuation and cosmic
density parameter at initial expansion parameter $a_i$ (\cite{P80},
\S 19E).  If $\Omega=\Omega_i=1$, the collapse of a spherical overdense
perturbation occurs at $a=a(\tau_c)\equiv a_c$ obeying the well-known relation
\begin{equation}
a_c^{-1}(\delta_0>0,\epsilon_0=0)=
{5\over3}\,\left(2\over3\pi\right)^{2/3}\,\delta_0=C_1\,\delta_0\ ,
\label{ac0}
\end{equation}
with $C_1=0.592954\ldots$.  In other words, if linear theory would imply
$\delta=1$ at $a=1$, then the density of a uniform spherical perturbation
becomes infinite at $a=C_1^{-1}=1.68647\ldots$.

Based on the above considerations, we are led to state the following
{\it Collapse Theorem} for overdense growing fluctuations:
A pressureless, irrotational mass element with initial density fluctuation
$\delta_i>(3/5)(\Omega_i^{-1}-1)$ collapses {\it at least as fast}
as a uniform spherical perturbation with the same initial $\delta$ and
$\dot\delta$, unless it first collides with another mass element.
The proof is trivial, following at once from equation (\ref{ddotdel})
combined with the solution of the spherical model.

Collision with another mass element inhibits collapse only if the matter
is collisional, in which case a shock wave is formed and the matter is
compressed a finite amount.  If the matter is collisionless, intersection
of mass elements always increases $\delta$ so that the collapse of each
individual stream is speeded up.

It is worth noting that the collapse theorem really makes two statements.
First, any growing-mode perturbation with $\delta_i>(3/5)(\Omega_i^{-1}-1)$
collapses (or collides with another element) in a finite time.  Second,
shear speeds up the collapse.  These two facts suggest two corollaries.
First, the collapse condition $\delta_i>(3/5)(\Omega_i^{-1}-1)$ may be too
weak in the presence of shear: initially underdense mass elements may
collapse.  Second, because the nonlinear density growth rate depends on
the shear, the initial shear affects the collapse time.  Therefore, initial
density maxima do not strictly identify the mass elements to collapse first.
To determine how shear actually affects the collapse it is necessary to
integrate all the evolution equations together, which we do in the next
section under the assumpation of vanishing $H_{ij}$.

The collapse theorem applies regardless of the spatial geometry of the
initial perturbation; it is based only on the trajectories of infinitesimally
nearby particles.  The theorem and its proof using the Raychaudhuri and
mass conservation equations is similar to proposition 4.4.1 of Hawking
\& Ellis (1973), which underlies one of Hawking's singularity theorems.
However, in the present case no spacetime singularity occurs (i.e.,
the spacetime remains geodesically complete) despite the collapse of
matter to infinite density because we have assumed that the fluctuations
are much smaller than the Hubble distance so that the Newtonian
gravitational potential (eq. [\ref{poisson}]), and therefore the metric
perturbations, remains small.

\section{General Solutions for $H_{ij}=0$}

We have shown that the nonlinear evolution of density perturbations
depends on the shear and tides.  To describe the evolution fully we
must integrate the equations of motion for these quantities in
addition to the density and expansion scalar.   We shall make the
simplifying assumption that the magnetic part of the Weyl tensor,
$H_{ij}$, may be neglected.

\subsection{Overdense Perturbations}

Nonzero shear causes overdense perturbations to collapse sooner than the
bound given by the collapse theorem.  To find out how much sooner, we have
integrated equations (\ref{lagcont}), (\ref{raych}), and (\ref{dsig})--(\ref
{dbet}) numerically for $\Omega=1$ with initial conditions given by equations
(16).  The expansion factor at collapse $a_c$ was evaluated as a function
of the initial tide parameters $(\epsilon_0,\alpha_0)$ for $\delta_0=1$.
Using equations (16) in the linear regime the results may be scaled to
any $\delta_0$.  Note that $\delta_0$ refers to the linear amplitude at
$a=1$; the actual integrations began at $a\ll1$.

The results are shown in Figure 1.  As expected, $a_c$ decreases with
increasing initial shear (or tide), verifying the collapse theorem.
Noticing that the contours are approximately elliptical, we fit the
following relation to our numerical results:
\begin{equation}
a_c^{-1}(\delta_0>0,\epsilon_0,\alpha_0)=C_1\,\delta_0+
C_2\,\epsilon_0\,\left(1-0.2\,\cos\alpha_0\right)\ ,\quad
C_2=1.25(1-C_1)\ .
\eqnum{19a}
\label{ac+1}
\end{equation}
The rms and maximum relative errors for $a_c$ in this fit are 1.5 and
3.6 percent, respectively, over the region shown in Figure 1.  The fit
fails to reproduce the asymmetry of the contours at large radii.  An
alternative fit, better for large $\epsilon_0$, is
\begin{equation}
a_c^{-1}(\delta_0>0,\epsilon_0,\alpha_0)=C_1\,\delta_0+
\epsilon_0\,\left(C_3+C_4\,s+C_5\,s^2+C_6\,s^3\right)
\eqnum{19b}
\label{ac+2}
\end{equation}
\setcounter{equation}{19}
where $s\equiv(1-\cos\alpha_0)^{1/3}$ and $C_3=0.36950$, $C_4=-0.04219$,
$C_5=0.26710$, and $C_6=-0.07404$.  The rms and maximum relative errors
are now 1.2 and 4 percent, respectively.

Plane-parallel perturbations, with $\epsilon_0=\delta_0$ and $\alpha_0=0$,
represent a special case of collapse with a simple exact solution to
our nonlinear equations: $\delta(\tau)=\delta_l/(1-\delta_l)$, $\theta=
\sigma=-\dot\delta_l/(1-\delta_l)$, $\epsilon=\delta_l$, and
$\alpha=\beta=0$, where $\delta_l(\tau)$ is the linear solution for
the growing mode.  This solution, valid for $\delta_l\le1$, is
identical to the result obtained by applying the standard Zel'dovich
(1970) approximation (\cite{MPS93}). Croudace et al. (1993) have
shown that the fully relativistic solution of this problem is given
by the Szekeres (1975) metric.

Zel'dovich (1970) suggested that, owing to kinematical effects, cold
gravitational collapse would generically approach the plane-parallel
solution.  Surprisingly, we find this conclusion to be invalid for the
solutions we have obtained.  As $\delta\to\infty$, the solutions generically
approach $\sigma=\theta\to-\infty$ as in the Zel'dovich solution but
with $\alpha+2\beta=3\pi$ and $\epsilon$ diverging (roughly as $\delta$)
rather than remaining finite.  Over much of the range of initial
parameters we find $\alpha\approx\pi$ at the moment of collapse,
corresponding to a velocity gradient tensor with one positive and
two negative (and equal) eigenvalues in the proportions $1:(-2+\alpha'):
(-2-\alpha')$ where $\alpha'\equiv(\pi-\alpha)/\sqrt{3}\ll1$.  In other
words, the final state is two-dimensional collapse into a spindle or
filament with expansion along the third axis, the opposite of what
Zel'dovich predicted.  This type of final configuration is achieved even
if the behavior in the linear and weakly nonlinear regime is nearly
one-dimensional, as we will see in \S 4.3 below.  The final state is
purely one-dimensional only if $\alpha_0=0$.

Croudace et al. (1993) noted the instability of the plane-parallel solution
and suggested that it was due to the neglect of the magnetic part of the
Weyl tensor.  This is a possibility we cannot disprove because we have
neglected $H_{ij}$.  However, we can determine why filaments form in
this limit: the prolate configuration is favored by the nonlinear tide-shear
coupling term in equation (7), which is stabilizing for one-dimensional
(pancake-like) configurations with $\alpha\approx\beta\approx0$ but
destabilizing for two-dimensional (filamentary) configurations with
$\alpha\approx\beta\approx\pi$.  For $\alpha=\beta=0$, $E_{ij}$ has
eigenvalues with signature $(-,-,+)$ and $\sigma_{ij}$ has ($+,+,-)$
(assuming $\epsilon>0$ and $\sigma<0$, the generic signs for collapse).
The eigenvalues of their product (after subtracting the trace in eq.
[\ref{tide}]]) have signature $(+,+,-)$, so that this term, opposite in
sign to $E_{ij}$, retards the growth of tides.  If $\alpha=\beta=\pi$,
all of the signatures are reversed except for the tide-shear coupling
term, which therefore becomes destabilizing.  This behavior is seen most
clearly in equation (\ref{deps}).  It is possible that the magnetic
term in equation (\ref{tide}) would, under some circumstances,
counter the tide-shear coupling.  However, this cannot be decided
without further investigation.  In any case, the results obtained
here warn one not to take the generality of the Zel'dovich pancakes
for granted given the neglect of nonlinear effects in the kinematical
treatment.

A complete physical explanation will have to await a Newtonian derivation
of equation (\ref{tide}), on which this result hangs.  However, the result
is not surprising when one recalls that the gravitational field remains
finite at a planar distribution of mass while it diverges at a linear
distribution.  While it is true that the divergence is greatest for a
point-like (spherical) configuration, this state is never reached unless
the shear vanishes.  As we have seen, shear accelerates collapse, so that
the spherical end state is disfavored.  Collapsing mass elements find the
next best solution, a prolate spindle.

It is now easy to understand the asymmetry of the contours in Figure 1.
Initially prolate configurations (with $\cos\alpha_0<0$) collapse sooner
than initially oblate configurations because the shear and tides are
already oriented in the most unstable configuration.

An important consequence of our result is that initial density maxima
do not correspond, in general, to minima of the collapse time.
Equations (19) show that the initial tide (or equivalently, shear) is of
comparable importance to the initial density in determining the collapse
time of a mass element.

\subsection{Underdense Perturbations}

Given the importance of shear and tides in accelerating the collapse of
overdense perturbations, it is natural to inquire whether they can cause the
collapse of underdense perturbations.  To answer this question we integrated
the equations of motion with $\delta_0=-1$.  As Figure 2 shows, we found
indeed that sufficiently large initial tide can turn an expanding
underdense perturbation into a collapsing overdense one.  Because the
collapse time can diverge (for sufficiently small initial tide), we graph
$a_c^{-1}(\delta_0=-1,\epsilon_0,\alpha)$.  This quantity vanishes at the
innermost contour; configurations with smaller tide never collapse before
trajectories intersect.  Including the dependence on $\delta_0$, the
numerical results are fit by
\begin{displaymath}
a_c^{-1}(\delta_0<0,\epsilon_0,\alpha_0)={3\over4}\,C_1(\epsilon_0+2\delta_0)
+\left(C_7-{1\over4}C_1\right)(\epsilon_0\cos\alpha_0+2\delta_0)+
\end{displaymath}
\begin{equation}
C_7\,\left[(\epsilon_0\cos\alpha_0+2\delta_0)^2+3\epsilon_0^2\sin^2
\alpha_0\right]^{1/2}\ ,
\label{ac-}
\end{equation}
with $C_7\approx0.0676$.  The rms and maximum absolute errors in this fit
for $a_c^{-1}$ are 0.021 and 0.046 over the range shown in Figure 2.
The fit reproduces the kink at $\epsilon_0=-2\delta_0$ and $\alpha_0=0$,
but the location of the $a_c^{-1}=0$ contour does not match the numerical
results exactly.  Note that a negative $a_c^{-1}$ implies that collapse does
not occur.

By comparing Figures 1 and 2 one may notice that for large $\epsilon_0$
the contours of constant $a_c$ are very similar for $\delta_0<0$ and
$\delta_0>0$.  In fact, excluding the kink region mentioned above,
equation (19b), provides an excellent fit to both cases! Not only can
negative density perturbations collapse in an Einstein-de Sitter
universe, with large initial tide ($\epsilon_0\gg\vert\delta_0\vert$)
they collapse at nearly the same rate as positive density perturbations.

If the initial tide is small enough then a perturbation does not
collapse before it intersects another mass element.  In this case the
mass drains out of the initial perturbation because its expansion rate
always exceeds the Hubble expansion rate ($\theta>0$).  At late times
these solutions approach the expanding spherical void solution
(Bertschinger 1985) with $1+\delta\propto\tau^{-3}$, $\theta={3\over2}\,
\dot a/a$, $\sigma\propto-\tau^{-4}\ln\tau$, and $\epsilon=\hbox{constant}$.
The velocity field becomes radial ($\sigma/\theta\to0$) despite the
nonspherical gravitational potential ($\epsilon/\delta$ remains finite)
because the net gravitational deceleration (proportional to $1+\delta$)
is so small that the perturbation becomes freely expanding, retaining
thereafter whatever shape it had prior to entering the homologous
expansion phase.

\subsection{General Perturbations}

It is worthwhile to unify the initial density and tide to obtain a
representation of the solutions that is valid for any initial condition.
We define a net perturbation $\Delta$ and an angle $\gamma$ so that the
linear density perturbation and tide are
\begin{equation}
\delta_0=\Delta_0\,\cos\gamma_0\ ,\quad \epsilon_0={\Delta_0\over\sqrt{5}}\,
\sin\gamma_0\ .
\label{delgam}
\end{equation}
The variables $(\Delta_0,\gamma_0,\alpha_0)$ are entirely equivalent
to the eigenvalues of $\nabla_i\nabla_j\phi$ (eq. [\ref{gradg}]).  The ranges
of the new variables are $\Delta_0\in[0,\infty)$ and $\gamma_0\in[0,\pi]$.
The square root factor is introduced to diagonalize the covariance matrix
of the variables $(\Delta_0,\gamma_0,\alpha_0)$ for a homogeneous and
isotropic random process.

If the linear density field $\delta_0(\vec x\,)$ is a gaussian random field,
applying the results of Doroshkevich (1970) we find the joint probability
distribution
\begin{equation}
dP(\Delta_0,\gamma_0,\alpha_0)=\exp\left(-{\Delta_0^2\over2\sigma_\delta^2}
\right)\,{\Delta_0^5\,\sin^4\gamma_0\,\sin\alpha_0\over6\pi\,
\sigma_\delta^6}\,d\Delta_0\,d\gamma_0\,d\alpha_0\ ,
\label{dprob}
\end{equation}
where $\sigma_\delta$ is the standard deviation of $\delta_0$.  Equation
(\ref{dprob}) is reminiscent of a three-dimensional gaussian distribution
in spherical coordinates aside from three extra powers of
$\Delta_0\sin\gamma_0$.  These factors arise because the gravity gradient
tensor is specified by three distinct diagonal entries of $E_{ij}$
(and angles giving the orientation of the eigenvectors which have been
integrated out) in addition to $\delta$.  As a result, the distribution
of $\Delta_0$ is shifted to larger values compared with a gaussian,
while $\cos\gamma_0$ is much more likely to be small than is
$\cos\alpha_0$, which is uniformly distributed in $[-1,1]$.  We therefore
change variables from $\gamma_0$ to
\begin{equation}
\eta_0\equiv1-{16\over3\pi}\int_0^{\gamma_0}\sin^4\gamma\,d\gamma=
1-{2\over\pi}\,\gamma_0+{4\over3\pi}\,\sin2\gamma_0-{1\over6\pi}\,
\sin4\gamma_0\ .
\label{eta}
\end{equation}
This variable is uniformly distributed (the measure $dP$ depends on
$\eta_0$ only through $d\eta_0$) for a homogeneous and isotropic
random process, ranging from $\eta_0=-1$ ($\gamma_0=\pi$) for tide-free
negative density perturbations to $\eta_0=+1$ ($\gamma_0=0$) for
tide-free positive density perturbations.  The distribution of the
remaining variable, $\Delta_0$, is a simple combination of gamma
distributions of the argument of the exponential in equation (\ref{dprob}).

Figure 3 shows the inverse collapse factor $(C_1\Delta_0\,a_c)^{-1}$ as
a function of the two variables $\cos\alpha_0$ and $\eta_0$, assuming
irrotational growing-mode perturbations in an Einstein-de Sitter universe.
Because these variables are independently uniformly distributed, this
diagram may be viewed as an equal-area representation of the general
case; all points in the plane are equally likely for homogeneous and
isotropic random initial conditions.  Using this interpretation and
computing the fraction of the area below the $a_{c}^{-1}=0$ contour,
we conclude that a randomly chosen mass element has a probability of
0.780 to collapse, implying that 56\% of the underdense perturbations
(and 100\% of the overdense ones) will collapse.  While the exact
numerical values depend on the assumption that $H_{ij}=0$, the
conclusion that some underdense regions must collapse is valid
in general because of the effects of shear.  Note that orbit-crossing
only hastens collapse since it increases the density and therefore
the gravitational focussing of trajectories.

The figure presents our results in a rather different way than Figures 1
and 2.  One's first impression may be that for a given $\Delta_0$ the
fastest collapse occurs for the most spherical initial perturbations.
However, there are two flaws in this conclusion.  First, although
increasing $\eta_0$ means increasing {\it ratio} of density to tide,
this does not contradict our earlier conclusion that for fixed $\delta_0$,
the slowest collapse occurs for spherical perturbations, because constant
$\delta_0$ and decreasing tide is not the same as constant $\Delta_0$
and increasing ratio of tide to density.  One must be careful to specify
what quantities are being held constant.  The second problem is that we
find a narrow region of large $\eta_0$ where the $(\Delta_0\,a_c)^{-1}$
exceeds the spherical collapse value $C_1$; Figure 3 shows the contour
$(C_1\Delta_0\,a_c)^{-1}=1.05$.  The contours double back at the top of
this region so that $(C_1\Delta_0\,a_c)^{-1}=1$ on the line $\eta_0=1$.

Although the collapse time appears to be a relatively simple function of
$\cos\alpha_0$ and $\eta_0$, we do not have yet an accurate  fitting formula.
Before investing much effort in this venture we should relax the assumption
$H_{ij}=0$, a task we leave for future work.

If we crudely approximate $(C_1\Delta_0\,a_c)^{-1}\approx{2\over3}\,
\eta_0+{2\over5}$, then local minima of the collapse time correspond to
local maxima of $\Delta_0({2\over3}\,\eta_0+{2\over5})$.  It would be
interesting to see in a realization of a gaussian random field how this
quantity differs from $\delta_0$, whose maxima are widely (and erroneously)
considered to be the first points to collapse.

Figure 4 shows the final shape of the collapsed object, as represented
by $\cos\alpha$, as a function of the initial parameters.  Recalling
that $\cos\alpha=-1$ for a two-dimensional (prolate) collapse while
$\cos\alpha=+1$ for a one-dimensional (oblate) collapse, we see that,
aside from regions with initially small shear ($\eta_0\approx\pm1$) and
oblate shape ($\cos\alpha_0\approx1$), most of the parameter space results
in collapses that are nearly two-dimensional.  In the shaded region the
eigenvalues of the velocity gradient tensor deviate by less than 10 percent
from the ratios $1:-2:-2$ for a filament collapsing along two dimensions.
This result confirms and extends what we found in \S 4.1 above.

\section{Conclusions}
In this paper we have developed and applied a Lagrangian fluid approach
to studying nonlinear gravitational instability.  This approach, pioneered
by Matarrese et al. (1993), has several advantages over the usual Eulerian
description on one hand and the Zel'dovich approximation on the other hand.
The first advantage is that we follow the fluid variables associated with
a given mass element, which is what we want to do if we are to track the
formation of objects by gravitational instability.  Second, the Lagrangian
fluid flow approach highlights the important physical role played by the
shear and tides, quantities that are not so apparent in the Eulerian
treatment and are generally neglected in the Zel'dovich approximation.

Going further requires making some assumption about the magnetic part
of the Weyl tensor $H_{ij}$, a quantity having no clear Newtonian
interpretation but which may, nevertheless, be present in the Newtonian
limit.  If it is negligible, as we assumed in \S 4, then the Lagrangian
approach has several additional advantages.  The first and clearest is
that the pressureless, irrotational flow problem reduces to a small set
of six ordinary differential equations for, effectively, the eigenvalues
of the velocity and gravity gradient tensors.  These equations are easy
to integrate to obtain accurate numerical solutions.  Second, the method
is exact (subject to the above-mentioned assumptions) until orbit-crossing.
The Zel'dovich approximation is exact for one-dimensional perturbations.
The Lagrangian method is exact for cylindrically and spherically symmetric
perturbations also (in these cases $H_{ij}$ vanishes by symmetry).
Third, this method is completely local in Lagrangian space (if $H_{ij}=0$)
until trajectories intersect, so that one can follow the evolution of
different mass elements independently up to orbit-crossing.  The Poisson
equation needs to be solved only once at the beginning to give the initial
tidal field.

These advantages come with a price: the Lagrangian fluid method does
not keep track of particle trajectories.  Thus, although one can compute
how the density of a given mass element changes with time, one cannot
know its position without separately integrating the equations of motion
for the trajectory.  The Lagrangian fluid equations could be integrated
together with the trajectories in order to supplement standard N-body
simulation methods.  Matarrese et al. (1993) have demonstrated a more
elegant method in which only differences in positions are integrated.
With standard N-body techniques one can compute the density and shear
only by averaging over volumes large enough to contain several particles.
With the fluid flow approach the density and other quantities would be
evolved for each particle, although the densities would have to be
coarse-grained in regions of multi-streaming.

In this paper we have stated and proven a theorem for pressureless,
irrotational self-gravitating flows: For a given initial $\delta$ and
$\dot\delta$, a uniform spherical perturbation (i.e., one with zero shear)
collapses (i.e., reaches infinite density) more slowly than any other
shape.  The proof requires no assumptions concerning the evolution of
shear or tides or the magnetic part of the Weyl tensor.  This theorem
applies to the evolution of individual mass elements rather than to the
global behavior of an extended mass distribution, for which spherical
symmetry may still be preferred as a state of lowest energy.  Moreover,
if one defines collapse by the requirement that all three axes collapse
for an ellipsoidal perturbation, then the spherical tophat collapses
most rapidly (White \& Silk 1979).  However, for many applications
one wants to know when density singularities first form, in which
case our collapse theorem is applicable.

The collapse theorem contradicts standard lore.  One's intuition suggests
that uniform spherical collapse ought to be most efficient because the
gravitational field of a point mass is stronger than that for linear or
planar distributions.  However, this intuition neglects the very important
role played by shear, which always acts to increase the rate of growth of
fluid convergence $-\vec\nabla\cdot\vec v$ following a given fluid element.
The effect of shear is not apparent in the Euler equations until one
writes the time derivatives following a given fluid element.

Although exact analytical results are not currently available, we have
numerically integrated the Lagrangian fluid equations assuming $H_{ij}=0$
to determine the actual collapse time for perturbations with shear and
to investigate the nature of the collapse.  We found three interesting
things.  First, underdense regions can collapse if the initial shear is
large enough.  One might think this to be unlikely, but we showed that
56\% of underdense perturbations (assuming a homogeneous and isotropic
random process) will collapse if not prevented by shell-crossing.
The exact percentage may differ if we relax the condition $H_{ij}=0$.

Second, local minima of the collapse time do not correspond to local
maxima of the initial density.  Instead, they correspond to maxima of
a combination of the eigenvalues of the initial gravity gradient (or
velocity gradient) tensor.  Although we have not obtained an analytical
formula for this linear combination, we have given a crude approximation
in terms of the variables $\Delta_0$ and $\eta_0$ defined in equations
(\ref{delgam}) and (\ref{eta}).

Third, we showed that when $H_{ij}=0$, the collapse is generically
two-dimensional, in contrast with the prediction of Zel'dovich (1970)
that gravitational collapse is one-dimensional.  Zel'dovich did not
consider the evolution of tides and he gave only a kinematical
extrapolation of the linear evolution of trajectories.  We find that,
even if the initial collapse is nearly one-dimensional, nonlinear
coupling between the tide and shear tends to drive the shape of the
collapsing mass into a filament rather than a pancake.  We gave a
heuristic explanation for this, noting that the gravitational field
of a filament is stronger than that of a pancake.  Although the field
of a point mass is still stronger, nonzero shear prevents uniform
spherical collapse.

Numerical simulations give conflicting answers to the question of
whether cold gravitational collapse is one- or two-dimensional.
Loeb \& Rasio (1994) find that shear can lead to a filamentary
collapse for irrotational perturbations, but White (1993, private
communication) found that a uniformly expanding triaxial ellipsoid
collapses along one dimension first, in good agreement with the
approximate analytic theory of White \& Silk (1979) and in disagreement
with the results obtained assuming $H_{ij}=0$.  It appears that $H_{ij}$
can cause a pancake-like collapse to occur in some circumstances.
Giving $H_{ij}$ a Newtonian interpretation and determining the conditions
under which it is significant are important tasks for the future.

Our results have several important implications for the formation of
galaxies and large-scale structure by gravitational instability.  First,
the widely-used spherical tophat collapse model underestimates the rate
of gravitational collapse.  A good deal of analytic theory in large scale
structure has been based on the spherical collapse model, including
the Press \& Schechter (1974) theory for the mass function of collapsed
objects.  In a numerical simulation Katz et al. (1993b) noticed that objects
with large initial shear collapse sooner as we would predict.  Hoffman
(1986) first made this point using the Zel'dovich approximation.  Second,
the two-dimensional character of at least some collapses means that
galaxies may be expected to form along filaments.  This point was also
noted by Katz et al.  Moreover, stronger initial tides for a given
overdensity imply stronger prolateness of collapsing dark halos, leading
to strong influences on the halo shape after virialization (\cite{D92}).
Third, initial density maxima are not the first objects to collapse.
Indeed, Katz, Quinn, \& Gelb (1993a) found from numerical simulations
that many of the first objects to collapse in a cold dark matter model
have no initial density maxima associated with them, calling into question
the widely-used peak-bias model for galaxy formation.  Fourth, matter in
underdense regions also collapses unless the initial shear is too low.
Thus, it is not clear that one expects galaxy formation to differ much
in voids and in moderate-density environments.

In future work we will investigate these implications.  We will also test
the results using N-body simulations and, conversely, test the N-body
method by its ability to reproduce our theoretical results.  A final
important area of study concerns the evolution of the Weyl tensor.
Although we have borrowed results from general relativity, the equations
of motion for $E_{ij}$ and $H_{ij}$ should be derivable from Newton's
laws.  Conversely, the Newtonian description should help to clarify
the proper relativistic treatment after orbits cross as well as the
interpretation of $H_{ij}$.

\acknowledgments

We wish to thank Andrew Hamilton, Rien van de Weygaert, Simon White,
and Fred Rasio for useful discussions.  This work was supported by
NSF grant AST90-01762.

\clearpage

\section{Figure Captions}

\noindent Fig.~1:
{Contours of constant collapse time, as indicated by the
cosmic expansion factor $a_c$, versus initial tidal field parameters
for linear density perturbation $\delta_0=+1$.  The light (heavy) contours
are spaced by 0.1 (0.5), with the central contour $a_c=1.6$ and the outermost
contour $a_c=0.4$.  The maximum collapse time, $a_c=1.686$, occurs at the
center corresponding to zero shear.}


\noindent Fig.~2:
{Contours of constant inverse collapse factor $a_c^{-1}$ for
initial negative density perturbations with $\delta_0=-1$.  The light
(heavy) contours are spaced by 0.1 (0.5), with the innermost contour
$a_c^{-1}=0$ and the outermost one $a_c^{-1}=1.8$.  Initial perturbations
in the central region do not collapse, while perturbations with $a_c^{-1}
>C_1=0.593$ ($a_c<1.686$) collapse faster than spherical {\it positive}
density perturbations of the same $\vert\delta_0\vert$.}


\noindent Fig.~3:
{Contours of constant inverse collapse factor $a_c^{-1}$,
normalized to the spherical value $C_1$, for perturbations with
$\Delta_0=1$ (see eq. [21] for the definition).  The shape of the
initial tidal tensor is specified by $\cos\alpha_0$, with $\cos\alpha_0
=-1$ ($+1$) for prolate (oblate) perturbations.  The parameter $\eta_0$
defines the ratio of density to tide, with $\eta_0=+1$ ($-1$) for
spherical positive (negative) density perturbations and $\eta_0=0$
for pure tidal perturbations with $\delta_0=0$.  The light (heavy)
contours are spaced by 0.1 (0.5), aside from the uppermost contour,
which is $a_c^{-1}=1.05\,C_1$.  The lowest contour is $a_c^{-1}=0$,
below which perturbations do not collapse.}


\noindent Fig.~4:
{Contours of constant final shear shape $\cos\alpha$ versus
the same initial tidal parameters shown in Figure 3.  Starting in the
upper right corner and moving to the lower left, the contours are 0
(heavy), $-.1$, $-.2$, $\ldots$, $-.9$ (heavy), $-.95$, $-.98$, $-.99$,
$-.999$ (heavy), and thereafter they reverse this sequence.  Over most of
this plane the final collapse is strongly prolate ($\cos\alpha\approx-1$).
The shaded region shows $\cos\alpha<-0.95$.}

\end{document}